\documentclass[aps, prl, superscriptaddress, reprint, floatfix, nobibnotes]{revtex4-2}
\usepackage{amsmath}
\usepackage{color}
\usepackage{comment}
\usepackage{bm,graphicx,hyperref}
\hypersetup{%
  breaklinks = {true},
  citecolor = {blue},
  colorlinks = {true},
  linkcolor = {red},}

\usepackage{lineno}
\usepackage[bottom]{footmisc}
\usepackage{gensymb}

\usepackage[dvipsnames]{xcolor}

\usepackage{titlesec}
\titleformat{\section}{\bf\huge\bfseries}{\thesection}{0pt}{}
\titleformat{\subsection}{\bf\large\bfseries}{\thesubsection}{0pt}{}

\titlespacing{\subsection}{0pt}{8pt plus 2pt minus 4pt}{0pt plus 0pt minus 0pt}
\begin{document}

\title{Atomically-precise Vacancy-assembled Quantum Antidots}
% AUTHORS AND AFFILIATIONS

\author{Hanyan Fang}
\thanks{These authors contributed equally to this work.}
\affiliation{Department of Chemistry, National University of Singapore, Singapore 117543, Singapore}

\author{Harshitra Mahalingam}
\thanks{These authors contributed equally to this work.}
\affiliation{Institute for Functional Intelligent Materials, National University of Singapore, Singapore 117544, Singapore}

\author{Xinzhe Li}
\thanks{These authors contributed equally to this work.}
\affiliation{School of Energy and Power Engineering, Xi'an Jiaotong University, Xi'an 710049, China}

\author{Xu Han}
\thanks{These authors contributed equally to this work.}
\affiliation{Department of Chemistry, National University of Singapore, Singapore 117543, Singapore}

\author{Zhizhan Qiu}
\affiliation{Institute for Functional Intelligent Materials, National University of Singapore, Singapore 117544, Singapore}

\author{Yixuan Han}
\affiliation{Department of Chemistry, National University of Singapore, Singapore 117543, Singapore}

\author{Keian Noori}
\affiliation{Institute for Functional Intelligent Materials, National University of Singapore, Singapore 117544, Singapore}
\affiliation{Centre for Advanced 2D Materials (CA2DM), National University of Singapore, Singapore 117546, Singapore}

\author{Dikshant Dulal}
\affiliation{Yale-NUS College, 16 College Avenue West, Singapore 138527, Singapore}

\author{Hongfei Chen}
\affiliation{Joint Key Laboratory of Ministry of Education, Institute of Applied Physics and Materials Engineering, University of Macau, Taipa, Macau 999078, China}

\author{Pin Lyu}
\affiliation{Department of Chemistry, National University of Singapore, Singapore 117543, Singapore}

\author{Tianhao Yang}
\affiliation{Department of Chemistry, National University of Singapore, Singapore 117543, Singapore}

\author{Jing Li}
\affiliation{Key Laboratory of Bio-inspired Smart Interfacial Science and Technology of Ministry of Education, School of Chemistry, Beihang University, Beijing 100191, China}

\author{Chenliang Su}
\affiliation{International Collaborative Laboratory of 2D Materials for Optoelectronics Science and Technology of Ministry of Education, Institute of Microscale Optoelectronics, Shenzhen University, Shenzhen 518060, China}

\author{Wei Chen}
\affiliation{Department of Chemistry, National University of Singapore, Singapore 117543, Singapore}

\affiliation{Centre for Advanced 2D Materials (CA2DM), National University of Singapore, Singapore 117546, Singapore}

\author{Yongqing Cai}
\affiliation{Joint Key Laboratory of Ministry of Education, Institute of Applied Physics and Materials Engineering, University of Macau, Taipa, Macau 999078, China}

\author{Antonio Castro H. Neto}
\affiliation{Institute for Functional Intelligent Materials, National University of Singapore, Singapore 117544, Singapore}
\affiliation{Centre for Advanced 2D Materials (CA2DM), National University of Singapore, Singapore 117546, Singapore}

\author{Kostya S. Novoselov}
\affiliation{Institute for Functional Intelligent Materials, National University of Singapore, Singapore 117544, Singapore}
\affiliation{Centre for Advanced 2D Materials (CA2DM), National University of Singapore, Singapore 117546, Singapore}

\author{Aleksandr Rodin}
\thanks{Corresponding author}
\affiliation{Centre for Advanced 2D Materials (CA2DM), National University of Singapore, Singapore 117546, Singapore}
\affiliation{Yale-NUS College, 16 College Avenue West, Singapore 138527, Singapore}
\affiliation{Materials Science and Engineering, National University of Singapore, 117575, Singapore}

\author{Jiong Lu}
\thanks{Corresponding author}
\affiliation{Department of Chemistry, National University of Singapore, Singapore 117543, Singapore}

\affiliation{Institute for Functional Intelligent Materials, National University of Singapore, Singapore 117544, Singapore}

\affiliation{Centre for Advanced 2D Materials (CA2DM), National University of Singapore, Singapore 117546, Singapore}

\date{\today}

\begin{abstract}

An antidot, defined in two-dimensional (2D) electron systems, is a region of a potential hill that repels electrons rather than attracts them, as in a quantum dot (QD). Patterning antidots (``voids") into well-defined antidot lattices creates an intriguing class of artificial structures for the periodic modulation of 2D electron systems, leading to anomalous transport properties and exotic quantum phenomena as well as enabling the precise bandgap engineering of 2D materials to address technological bottleneck issues~\citep{Weiss1991,Fleischmann1992,Gunawan2008,Ouyang2011,Cupo2017,Du2018,Liu2021}. Nanolithography through encapsulated layer bring the conventional antidots from semi-classcial regime to quantum regime~\citep{Sandner2015,Jessen2019}. To advance further, it is crucial to have precise control over the size of each antidot and its spatial period at the atomic scale, which enables the transition to a new paradigm by creating quantum antidots (QADs) hosting discrete quantum hole states as promising candidates for single-photon emitters~\citep{Du2009,Mitterreiter2021}, spin Qubits~\citep{Flindt2005,Pedersen2008}, plasmonic photocatalysis~\citep{Besteiro2017} and also for exploring
collective dynamics of interacting quasiparticles and hot electrons in previously inaccessible regimes~\citep{Zhang2013,Clavero2014}. However, realizing such atomic-scale QADs is infeasible by current top-down techniques or through engineering QAD behaviors in quantum Hall regime~\citep{Goldman1995,Maasilta2000,Sim2003} because the latter would require an enormous and inaccessible magnetic field up to hundreds of Tesla. Here, we report an atomically-precise bottom-up fabrication of a series of atomic-scale QADs with size-engineered quantum states through a controllable assembly of a chalcogenide single vacancy (SV) in 2D PtTe$_2$, a type-II Dirac semimetal~\citep{Yan2017}. Te SVs as atomic-scale ``antidots" undergo thermal migration and assembly into highly-ordered SV lattices spaced by a single Te atom, reaching the ultimate downscaling limit of antidot lattices. Increasing the number of SVs in QADs strengthens the cumulative repulsive potential and consequently enhances collective interference of multiple-pocket scattered quasiparticles inside QADs, creating multi-level quantum hole states with tunable gap from telecom to far-infrared regime. The inter- and intra-valley scattering around the anisotropic $M$-pocket is found to be the dominant source of electronic scattering, realizing extraordinary fine-tuning of the nodal patterns of these quantum states. Moreover, precisely engineered qunatum hole states of QADs are symmetry-protected and thus survive upon oxgygen substitutional doping. Therefore, SV-assembled QADs exhibit unprecedented robustness and property tunability, which not only hold the key to their future applications but also embody a wide variety of material technologies, including introducing hetero-atoms with various potential centers to create quantum hetero-antidots and introducing spin-polarized atoms to create magnetic QADs for potential applications in spintronics.

\end{abstract}	
\maketitle

Atomic-precision manufacturing of artificial quantum matter with designer quantum states towards atomic-scale devices is central to quantum nanotechnology~\citep{Fuechsle2012,Huff2018,Achal2018,Kalff2016}. Such a representative device constitutes a zero-dimensional (0D) quantum dot with attractive potential from a finite atomic ensemble~\citep{Amlani1999,Imre2006,Kim2014,Folsch2014}, which confines electrons into discrete 0D quantum states. A complete opposite of a QD, quantum antidot (QAD) with a potential hill confines a finite droplet of holes in discrete quantum states, which can be promising candidates for single-photo emitters or spin Qubits. In addition, the operation of such devices under a magnetic field where competing length scales harbor rich physics enables the manipulation of exotic charge transport, spin textures, and collective quasiparticles dynamics with nontrivial topological properties in both integer and fractional quantum Hall regime~\citep{Goldman1995,Maasilta2000,Sim2003}. As the trend towards miniaturized devices continues, creating atomically precise microscopic QADs with high digital fidelity and resistance to environmental perturbations is a game changer, indispensable for a broad range of emergent quantum technologies. 

The advances in nanolithographic techniques make it possible to impress antidot and their lattice of different sizes, shapes, and periods in 2D systems from semi-classic regime~\citep{Park1997,Sinitskii2010} to quantum regime~\citep{Sandner2015,Jessen2019}. Nanopatterning on hexagonal boron nitride-encapsulated graphene reduces the antidot diameter and lattice period, resulting in high mobility and reduced edge disorder, leading to pronounced commensurability features~\citep{Sandner2015,Jessen2019}.  However, QAD generated by nanolithography still exhibits inherent atomic structure randomness and thus insufficient digital fidelity. Moreover, the complete voids created by nanolithography cannot host quantum confined states in the empty void but only leave the localized edge states. To further reduce the antidot size to the ultimate downscaling limit, scanning tunnelling microscope provides a precise control in designing artificial electronic and spin lattices via atomic manipulation~\citep{Khajetoorians2019}. Molecular graphene with Dirac-like linear dispersion has been created by assembling carbon monoxide (CO) molecules into a triangular lattice on copper surface, where honeycomb geometry in 2D electron gas is formed by this anti-framework~\citep{Gomes2012}. The flexibility of the CO arrangements also enables the realization of other artificial quantum structures such as Lieb~\citep{Slot2017} and breathing Kagome~\citep{Kempkes2019a} lattices with tailored electronic structure but weak thermal stability. Alternatively, creating vacancy lattices in a chlorine-terminated Cu (100) produces robust quantum structures that can survive at 77K~\citep{Kalff2016}, which can be utilized to design topological domain-wall states and Lieb lattice with a flat band~\citep{Drost2017}. In contrast, realizing atomic-scale QAD behaviors in quantum Hall regime via magnetic field would require a gigantic magnetic field up to hundreds of Tesla, far beyond the strongest magnetic field available in a laboratory setting as the magnetic length (defined by $l_b =\sqrt {\frac {\hbar c}{eB}} \approx 26$ nm $/ \sqrt {B}$) competes with geometric length. Although engineering vacancies into void clusters has also been proposed in 2D and 3D systems ~\citep{Xu2002,Liu2014,Nguyen2018}, assembly of atomic vacancies into atomically-precise QADs remains infeasible to date, as one must tackle multiple complexities including the ability to achieve the precise vacancy superlattice formation together with the strong repulsive perturbation of quasiparticles with the short wavelength nature in the host. 

\begin{figure*}[t]
    \centering
    \includegraphics[width=\textwidth]{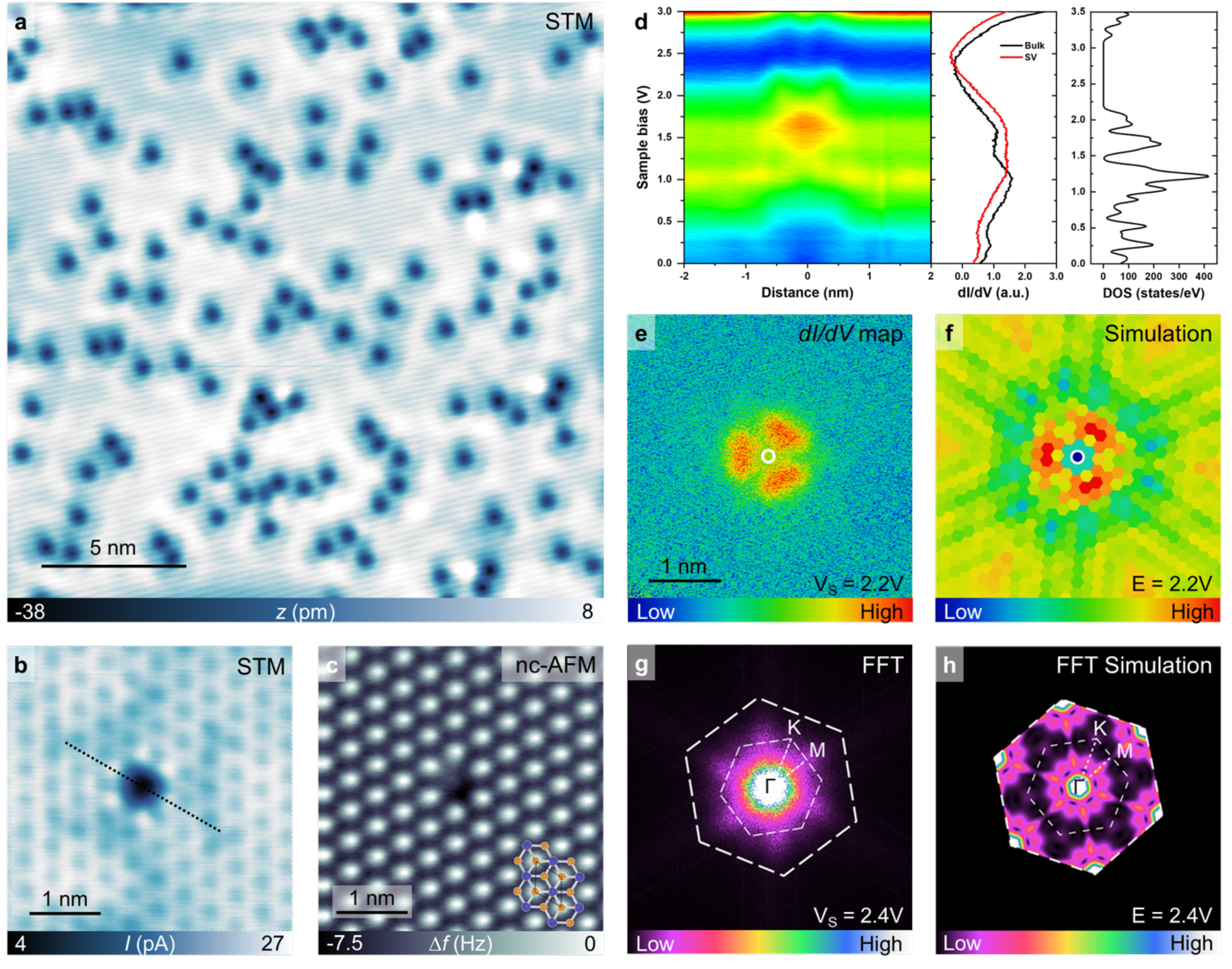}
    \caption{\textbf{$\vert$ Characterization of single Te vacancy on PtTe$_2$ surface.} \textbf{a}, Large scale STM image of PtTe$_2$ surface with a high density of single Te vacancies. ($V_S = -2.5$V). \textbf{b}, Constant height STM image and \textbf{c}, nc-AFM image for an isolated single Te vacancy ($V_S = 1$mV). \textbf{d}, Left panel: color-coded $dI/dV$ spectra taken along the line across over SV (marked by black dashed lines in \textbf{b}). Center panel: point $dI/dV$ spectra taken above SV and at the pristine surface. Right panel: the calculated DOS for bulk PtTe$_2$. \textbf{e}, $dI/dV$ map taken for SV at $V_S = 2.2$V at BB continuum. \textbf{f}, Simulated scattering pattern for SV at $V_S = 2.2$V. \textbf{g}, FFT result calculated from the $dI/dV$ map taken for SV at $V_S = 2.4$V inside the gapped region. \textbf{h}, Simulated FFT map originated from the scattering at BB.}
    \label{fig:SV}
\end{figure*}

Here, we demonstrate that in a 2D three-atom-layer transition metal dichalcogenide (TMDs), a single chalcogenide vacancy in platinum telluride (PtTe$_2$), can act as an atomic Lego for the atomically-precise bottom-up fabrication of a series of atomic-scale QADs. A mild diffusion barrier of Te SV in PtTe$_2$ ensures the feasibility of vacancy migration and assembly into periodic superlattices, wherein adjacent SVs are spaced at a single atom limit. Periodic arrangement of SV lattice in a microscopic region of PtTe$_2$ can be viewed as a partial removal of SV with atomic-scale spacing, enabling the formation of geometrically well-defined QADs with both edge-localised states and precisely engineered quantum states inside the QADs. Assembly of SVs in such a configuration in one-atom-layer or two-atom-layer 2D materials tends to collapse into a complete void due to a lack of support. Despite the semimetallic nature of PtTe$_2$, SV with a repulsive potential acts as an atomic-scale ``antidot", yielding strong perturbation to the short-range quasiparticles near the high-energy gapped regime, creating multi-level quantum hole states with tunable energies and nodal patterns. Joint DFT and quantum field theoretical calculations uncover that the inter- and intra-valley scattering at the $M$-pocket dominates the quasiparticles scattering, enabling the modulation of the nodal pattern of these bound states at the extraordinarily fine scale, unobtainable by conventional methods. Moreover, vacancy-based QADs host symmetry-protected bound states and exhibit unprecedented environmental robustness and property tunability upon atom-by-atom doping.

\subsection{Single vacancy as atomic Lego}

The density of Te SVs in bulk PtTe$_2$ crystals can be controlled by adjusting the growth conditions. The extended post-growth annealing at 1150$^{\circ}$C for 1h results in a high density of Te SVs ($3.1\pm 0.3\times 10^{13}$cm$^{-2}$), as manifested by dark spots in the large-scale STM image (Fig.~\ref{fig:SV}a), which provides sufficient building blocks for the subsequent assembly into QADs. This is also inspired by our previous study, revealing that Te SVs in PtTe$_2$ tends to cluster upon thermal annealing~\citep{LiXZ2021}. The close-up STM image taken over the SV at small sample bias $V_S = 1$ mV (Fig.~\ref{fig:SV}b) reveals an atomic void surrounded by the regular hexagonal lattice with a periodicity of $0.40 \pm 0.02$ nm, consistent with a previous report~\citep{Zhussupbekov2021}. We then performed constant-height nc-AFM imaging with a CO-functionalized tip to further confirm the atomic structure of SV by recording the spatial variation of frequency shift ($df$) in the Pauli repulsive regime~\citep{Leo2009}. The nc-AFM image (Fig.~\ref{fig:SV}c) unambiguously resolves the periodic arrangement of Te atoms on the surface except for the one missing Te atom in the center of defect site. However, the oxygen~\citep{Barja2019} or carbon~\citep{Schuler2019ACSnano,Cochrane2021} substituted chalcogen vacancy may yield a similar nc-AFM contrast. A detailed DFT calculation of electronic structures of several proposed structures, along with the simulation of nc-AFM image, and comparison with experimental data, allows us to exclude these possibilities (Extended Data Fig. 4 and 8).

To decipher the local electronic structure, bias-dependent line $dI/dV$ spectra are taken across the SV (indicated by a black dashed line in Fig.~\ref{fig:SV}b), as shown in the left panel of Fig.~\ref{fig:SV}d ($0$ nm refers to the center position of SV). Non-zero local density of states (LDOS) near the Fermi level ($E_F$) reflects the semimetal nature of bulk PtTe$_2$. Two broad peaks at $1.00\pm 0.02$V and $1.56\pm 0.02$ are mainly contributed from the Te contributed bands (Extended Data Fig. 5), consistent with previous report~\citep{Guo1986}. In addition, the presence of a broad dip feature from 2.15 eV to 2.70 eV suggests a significantly reduced LDOS, in good agreement with the perdicted bandgap from $\sim 2.2$ eV to 3.0 eV above $E_F$ (Fig.~\ref{fig:SV}d). To streamline the discussion, we denote the upper band above this gap as ``top band'' (``TB'') and the lower band as ``bottom band'' (``BB''), with the band minimum and maximum of the ``TB'' and ``BB'' denoted as ``TBM'' and ``BBM'', respectively. We observed an accumulation of LDOS near BBM over the SV, together with an upward band bending that extends 1 nm away from the SV center (details in Extended Data Fig. 3). This suggests that SV is negatively charged with a local repulsive potential ~\citep{Aghajanian2020,Fang2022}, thus bending the states close to BBM upwards into the gap. The $dI/dV$ map (Fig.~\ref{fig:SV}e) taken at energies near BBM reveals a triangular clover pattern localized within $0.40\pm 0.02$ nm away from the vacancy center, presumably arising from the scattering of quasiparticles from the negatively charged SV. To further understand the scattering at SV inside the gap, we performed fast fourier transform (FFT) to obtain the quantum interference pattern (Fig.~\ref{fig:SV}g). The intensity spots are observed around the $M$-point along the $\Gamma$-$M$ direction, which indicates the intra- and inter-valley scattering between $M$-pockets is dominant over the SV. The inter-valley scattering results in much shorter wavelengths compared to intravalley scattering, consistent with the $dI/dV$ map.

\begin{figure*}[t]
    \centering
    \includegraphics[width=\textwidth]{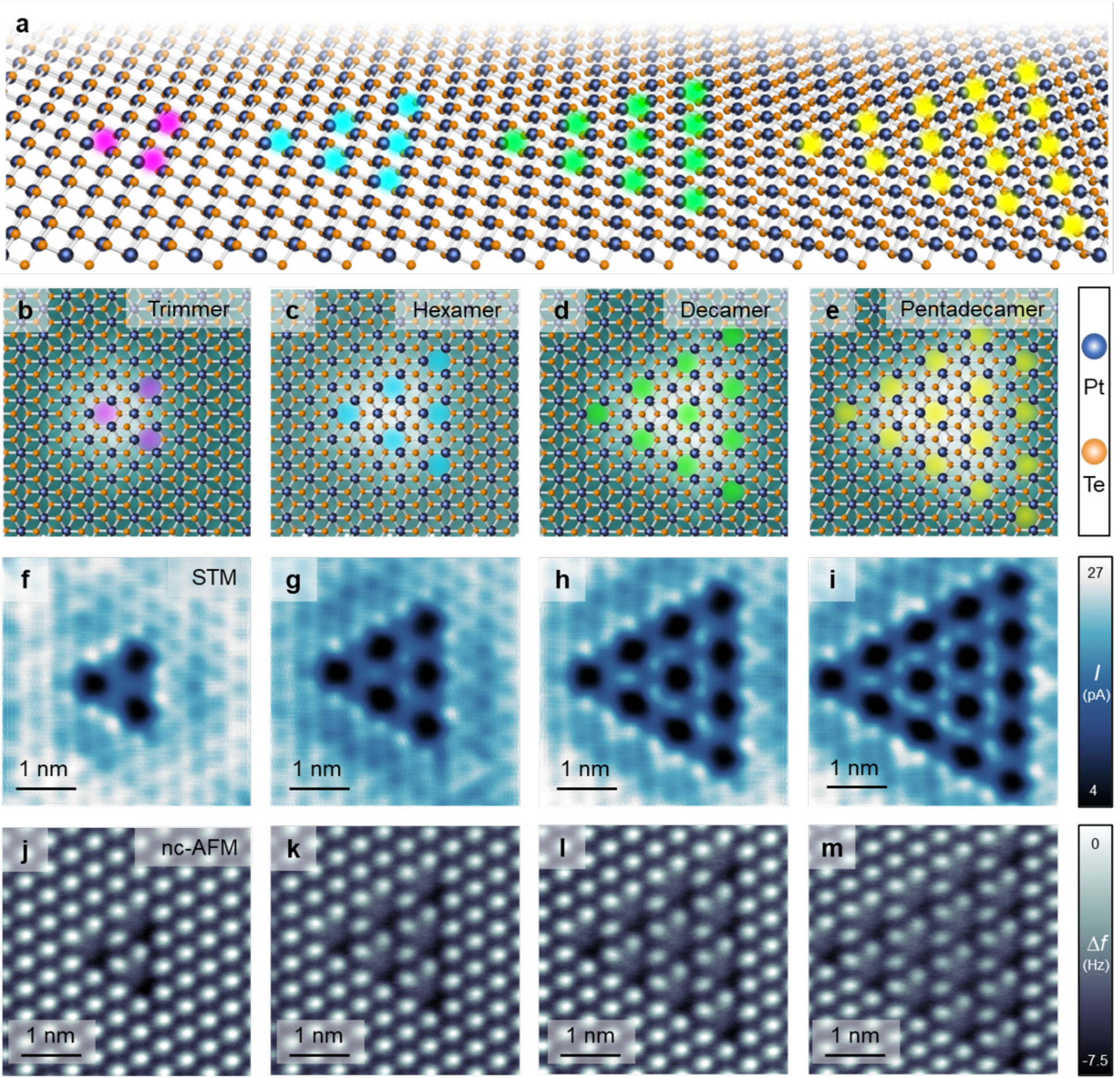}
    \caption{\textbf{$\vert$ Atomically-precise QADs assembled by a single Te vacancy superlattice on PtTe$_2$ surface.} \textbf{a}, Schematics highlight the bottom-up self-assembly of a series of triangular QADs on PtTe$_2$ surface. \textbf{b-e}, Atomic structure \textbf{f-i}, Constant height STM images and \textbf{j-m}, nc-AFM images of geometrically well-defined triangular QADs ranging from trimer to hexamer to decamer to pentadecamer. Set point: $V_S = 1$ mV. }
    \label{fig:QAD}
\end{figure*}

\subsection{SV-by-SV assembly of QADs}

Upon the thermal annealing above 433 K, we observed the appearance of atomically-precise triangular QADs on PtTe$_2$ surface accompanied by a decreased density of isolated SVs after each annealing process (Fig.~\ref{fig:QAD}). A statistical analysis further reveals that the total number of atomic voids remains approximately constant before and after thermal annealing. This suggests that such a mild annealing below 473 K mainly drives the migration and self-assembly of SVs into regular QADs rather than generating more Te vacancies on the surface, consistent with previous report that no chalcogen vacancy was generated after mild annealing at 250 $^{\circ}$C in WS$_2$~\citep{Schuler2019PRL}. Therefore, thermal annealing is likely to trigger the migration and assembly of the Te vacancy on the topmost layer into regular QADs. While achieving a mono-dispersive distribution of QADs with identical patterns remains a challenge, we have observed that the majority size and shape of QADs can be steered by adjusting the annealing temperature and duration (Extended Data Fig. 1). As illustrated in Fig.~\ref{fig:QAD}a, Te SVs can form triangular QADs with increasing lateral size. All the neighboring Te vacancies inside these QADs (Fig.~\ref{fig:QAD}b-e) are separated by twice of the lattice constant but leave one Te atom between two adjacent SVs to form a periodic superlattice. Assembly of 3, 6, 10, 15 SVs creates a set of atomically-precise QADs, denoted as trimer, hexamer, decamer and pentadecamer respectively, which can be resolved by STM (Fig.~\ref{fig:QAD}f-i). The detailed atomic arrangement of Te atoms and Te vacancies in these QADs can be directly visualized by nc-AFM (Fig.~\ref{fig:QAD}j-m). In addition, we also applied Kelvin probe force microscopy (KPFM) to characterize the local contact potential difference (LCPD) for all the assembled QADs with increasing size. Extended Data Fig. 6 presents the line KPFM results acquired along the edges of different QADs at a fixed tip sample distance. A parabolic fitting of bias-dependent frequency shift curves ($df-V$) reveals a positive shift of LCPD (i.e parabolic maximum) from $-234$ mV (SV) to $-161$ mV (trimer) to $-96$ mV (hexamer) to $-43$ mV (decamer), as the size of QADs grows. A smaller negative LCPD value over a larger QAD is associated with an increased collective repulsive potential, which may originate from negative charge accumulation or an averaging effect ~\citep{Gross2014}.

\begin{figure*}[t]
    \centering
    \includegraphics[width=0.9\textwidth]{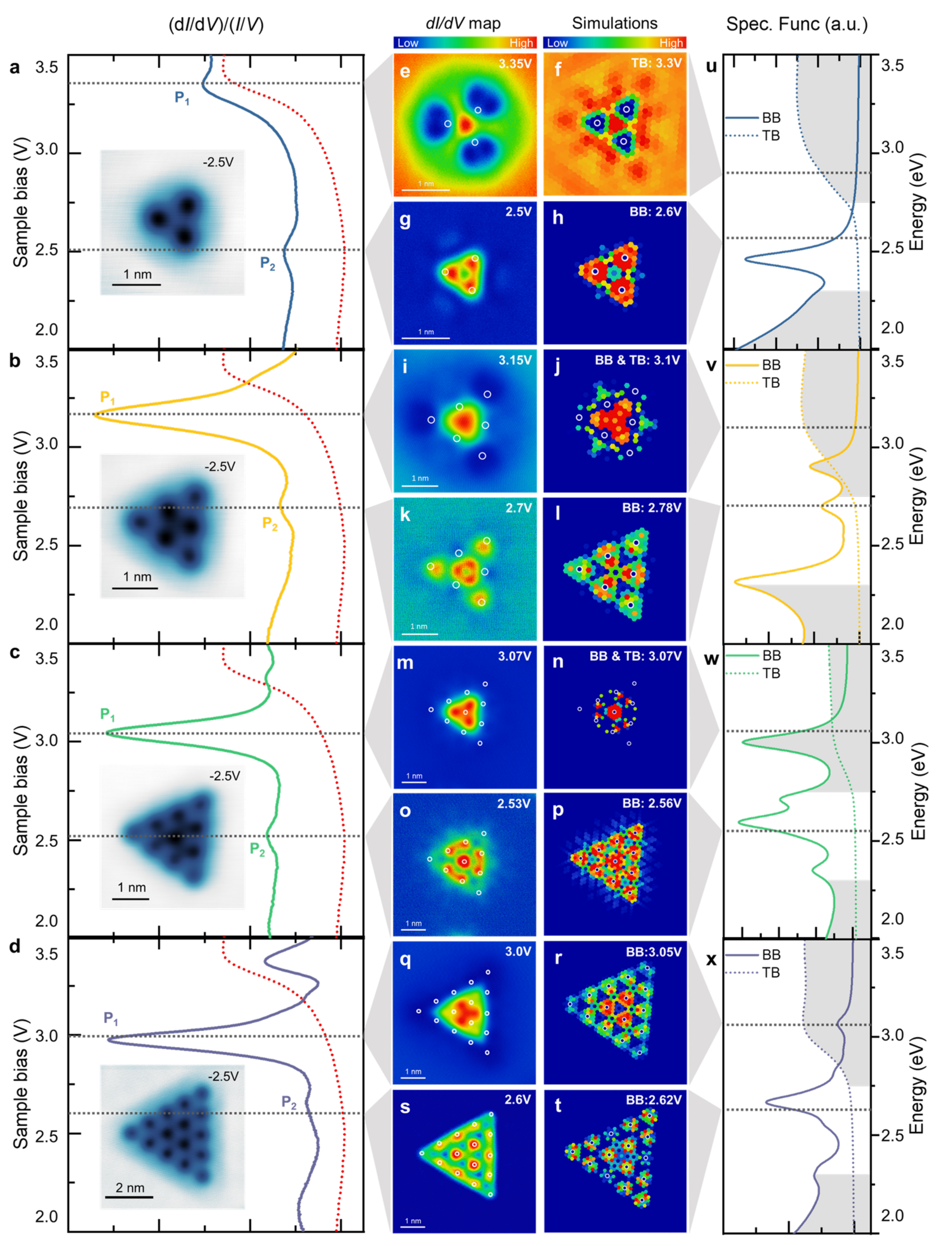}
    \caption{\textbf{$\vert$ Electronic characterization of atomically-precise triangular QADs with molecule-like quantum states.} \textbf{a-d}, STS taken at the center of trimer, hexamer, decamer and pentadecamer with reference curve of STS taken at SV in red dashed line. The corresponding STM images of the QADs are shown in the inset. Comparisons between experimental $dI/dV$ map and simulated spectral function maps from the field-theoretical scattering model are shown for \textbf{e-h}, trimer, \textbf{i-l}, hexamer, \textbf{m-p}, decamer, \textbf{q-t}, pentadecamer, where the energy positions are marked by dashed lines in \textbf{a-d}. \textbf{u-x}, Calculated spectral function from scattering model. The BB and TB indicate the spectral functions from the BB and TB bands(marked by grey shadow), respectively.}
    \label{fig:QuantumStates}
\end{figure*}

\subsection{Precisely engineered quantum hole states}

We then performed $dI/dV$ measurements to probe the size-dependent electronic properties of QADs. In addition to the enhanced upward band bending, normalized $dI/dV$ spectra taken at the center of QADs from trimer to pentadecamer reveal a set of new quantum states that vary with the size (Fig.~\ref{fig:QuantumStates} (a-d)). In contrast to SV (red curve), the trimer exhibits two resonance peaks at $3.35 \pm 0.03$ V (labeled as P$_1$) and $2.50 \pm 0.05$ V (labeled as P$_2$), whereas a pronounced peak at $3.17 \pm 0.02$ V, $3.05 \pm 0.02$ V, $2.97 \pm 0.02$ V (P$_1$) together with a weak peak at $2.70 \pm 0.08$ V, $2.53 \pm 0.06$ V, $2.65 \pm 0.07$ V (P$_2$) are observed for hexamer, decamer and pentadecamer, respectively. Both energetic positions of P$_1$ and the peak spacing between P$_1$ and P$_2$ decreases when the size of QAD increases. Additional features above P$_1$ can be observed for larger QADs such as decamer and pentadecamer since the energy spacing between emerged peaks reduces. These observations indicate that QADs in PtTe$_2$ generate multiple quantum states with size dependence. The intensity of these peaks also shows spatial variation within all QADs. In addition, we also observed extra bound state with prominent intensity localized in the particular region at the edge (Extended Data Fig. 7). Here, we first focus on the most intriguing bound states that fall inside the gapped region (P$_1$ near TBM and P$_2$ near BBM). $dI/dV$ mapping at the selected energies (marked by black dash lines in Fig.~\ref{fig:QuantumStates} (a-d)) reveals the corresponding spatial shape of these bound states for all the QADs. The mappings at P$_1$ show the accumulation of the LDOS at the center region for all the QADs. Additional fine structures at the center for the larger decamer and pentadecamer are due to electronic modulation from SVs at the center. The mappings at P$_2$ reveal more nodal patterns than P$_1$ for all the QADs . Particularly, small-sized QADs (trimer and hexamer) show an enhanced LDOS at the corners, while larger-sized QADs (decamer and pentadecamer) reveal a higher LDOS at the central vacancies and edges. 

\begin{figure*}[t]
    \centering
    \includegraphics[width=\textwidth]{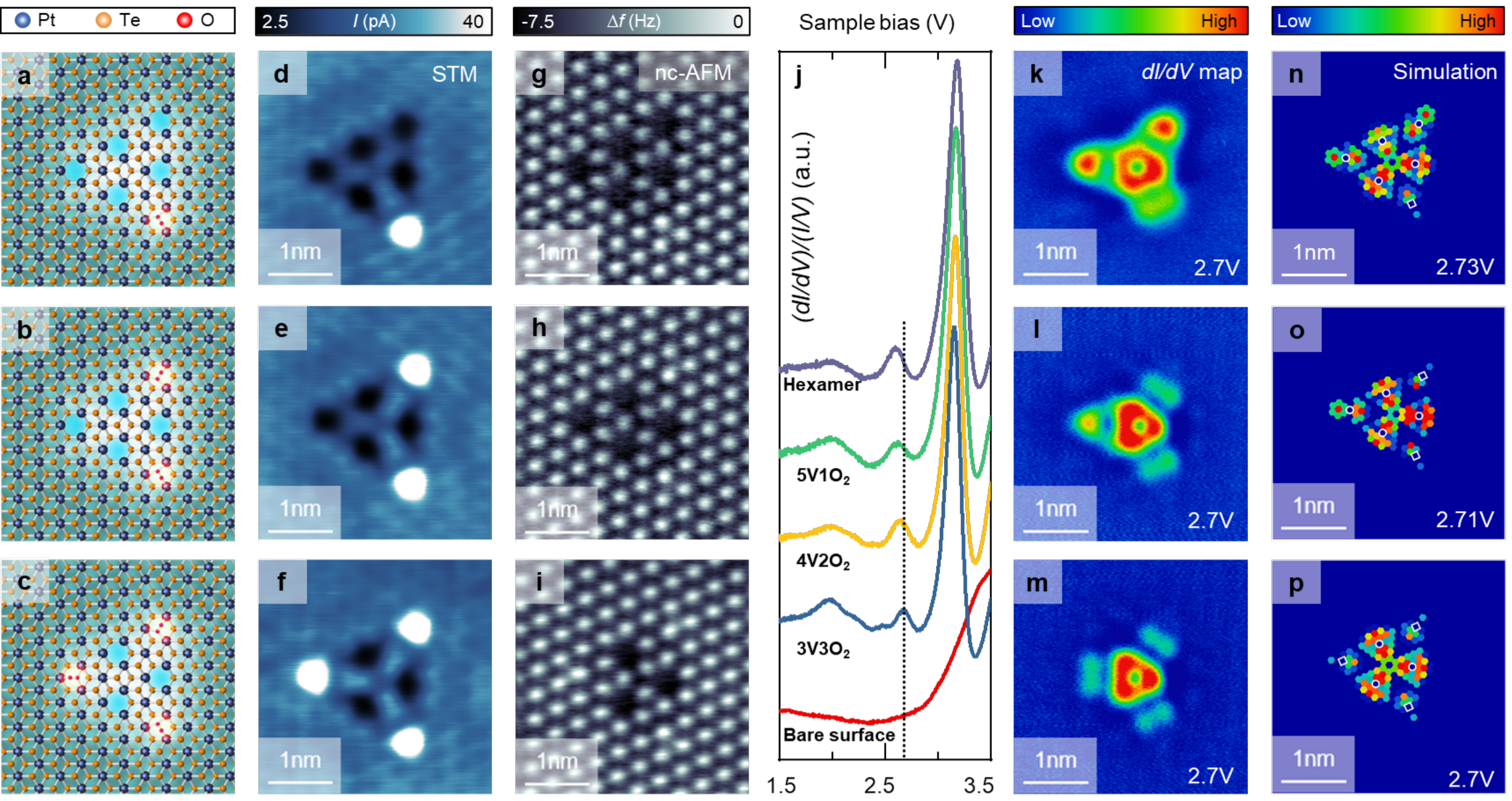}
    \caption{\textbf{$\vert$ Atom-by-atom oxygen substitution of Te SVs in QADs.} \textbf{a-c}, Schematic illustration of the atomic structure for a hexamer with 1 (5V1O$_2$), 2 (4V2O$_2$) and 3 (3V3O$_2$) corners being substituted by oxygen. The blue, yellow and red circles represent Pt, Te and oxygen atoms. \textbf{d-f}, Constant height STM images ($V_S = 1$ mV) and \textbf{g-i}, nc-AFM images for the oxygen-doped hexamer including 5V1O$_2$, 4V2O$_2$ and 3V3O$_2$. \textbf{j}, STS taken at the center of hexamer (6V) and oxygen-doped hexamer (5V1O$_2$, 4V2O$_2$, 3V3O$_2$) with the reference curve. The dashed line marks the energy position of bound state for undoped hexamer as a reference. \textbf{k-m}, $dI/dV$ maps taken at the peak energy in \textbf{j}, for the oxygen-doped hexamer and \textbf{n-p}, corresponding simulated spectral function maps by taking the O substitute as a weaker scattering center.}
    \label{fig:Oxygen}
\end{figure*}
\subsection{Atom-by-atom oxygen substitution}

Since chalcogen vacancies can be readily substituted by oxygen for TMDs material after exposure to air~\citep{Barja2019}, we evaluated the stability of these QADs and the impact of the oxygen on the QADs, crucial for their practical applications. After exposing the sample to air or dosing the oxygen to the UHV chamber at room temperature and maintaining the pressure at $1\times 10^{-7}$ mbar for 10 minutes, the majority of these QADs still remain on the surface, as evidenced by large-scale STM images. A close-up examination (Extended Data Fig. 8) reveals extra bright protrusions appearing at the corners of a small amount of QADs. Our DFT calculations indicate that a small energy barrier of 0.31 eV is required to transit from the adsorption of the O$_2$ molecule to the most stable state, in which two O-O atoms bond with three adjacent undercoordinated Pt atoms in the SV. This configuration is energetically favored by 0.22 eV compared to single-O-atom substitution (Extended Data Fig. 8). Taking hexamers as one example, we have identified three types of oxygen-substituted hexamers, depending on number of doped oxygen molecules. (Fig.~\ref{fig:Oxygen}(a-c)). Close-up STM images (Fig.~\ref{fig:Oxygen}(d-f)) show three hexamers with an increased number of protrusions at the corners, while nc-AFM image (Fig.~\ref{fig:Oxygen}(g-i)) reveals that atomic void is filled up by extra atoms at the corners with the enhanced STM contrast. We used the format of $\mathbf{xVyO_2}$ to describe the number of vacancies ($\mathbf{x}$) and O-O substitutes ($\mathbf{y}$) in each QAD. Here, we took the oxygen-substituted hexamer as a representative example to probe the impact from oxygen on local electronic structures. Although 3V3O$_2$ only has three vacancies, which form a trimer at the center, the $dI/dV$ spectra taken at center, edge and corner of 3V3O$_2$ remain almost unaltered to the hexamer but is very different from that of the trimer (Extended Data Fig. 9). From the point STS taken at the center (Fig.~\ref{fig:Oxygen}j) of hexamer with different number of O-O substitutes, we observe a slight energy shift of the quantum states as an increase of oxygen substitution. The  $dI/dV$ maps taken at $P_1$ and $P_2$ (Fig.~\ref{fig:Oxygen}(k-m)) for 5V1O$_2$, 4V2O$_2$ and 3V3O$_2$ preserve rather similar spatial LDOS distributions as the hexamer in the central region but only show a slightly reduced intensity  at the corners. These results suggest that the quantum states of these QADs are defined by the spatial geometry of potential perturbation and are robust against the oxygen substitution/doping.

\subsection{Discussion}

PtTe$_2$ is a TMD with a relatively low vacancy migration barrier, enabling the formation of regular vacancy clusters through a bottom-up self-assembly method. When a single vacancy migrates to its neighboring Te site, the corresponding Te atom must move in the opposite direction. To examine the possible diffusion pathways for moving the neighboring Te atom to the vacancy site, we use Climbing Image Nudged Elastic Band (CI-NEB) method (Extended Data Fig. 2). The most energy-favorable pathway involves the Te atom breaking two Pt-Te bonds and rotating around the Pt atom, while preserving the remaining Pt-Te bond and swinging over the Te atom at the bottom. The Te atom then fills in the vacancy site and bonds with Pt atoms again, completing the whole process. Our calculations reveal a relatively low diffusion barrier (1.61 eV) of a single Te vacancy in PtTe$_2$, much lower than the case in MoS$_2$ (2.27 eV) and other TMDs~\citep{Liu2014,Cai2016}. The diffusion barrier can be further reduced to 1.44 eV if SV is negatively charged. Such a mild barrier can be readily overcome to trigger the diffusion and assembly of SVs in PtTe$_2$ by annealing at 393K. The presence of vacancy leads to local lattice contraction and results in attractive interaction ~\citep{Trevethan2014}. Therefore, the identical separation of twice the lattice constant is likely a consequence of competition between vacancy-induced strain field and Coulomb repulsion between negatively charged SVs. Therefore, Te SV serves as an ideal building block for the construction of atomically-precise QADs. By controlling the thermal treatment temperature and duration, we can control the dominant type of QADs on the surface in size and shape (Fig.~\ref{fig:QAD}a).  

To decipher the physical origin of these quantum states in vacancy-based QADs, we first gained a deeper understanding of the electronic structure of the basic building unit, individual SVs. Since the negatively charged SV in PtTe$_2$(Extended Data Fig. 3) acts as a repulsive potential perturbation to electrons in the host, we can describe such local potential variation using an electron scattering framework that employs the nearly-free electron model to describe the surface states of the PtTe$_2$. For the BB, the dominant source of electron scattering close to the bandgap is the energy pocket centered around the $M$-point (Extended Data Fig. 5). Due to the symmetry of $M$-point in the hexagonal Brillouin Zone (BZ) of the lattice, this results in both inter- and intra-valley scattering from the surface-state electrons. For the TB, the $\Gamma$-point and $K$-point pockets are the dominant source of scattering close to the bandgap. For both bands, we approximate the dispersion around the high-symmetry points with a 2D asymmetric parabolic dispersion, with direction-dependent carrier effective masses. Indeed, the calculated spatial FFT map of the SV (Fig.~\ref{fig:SV}g) exhibits the signature of intervalley scattering between the six $M$ points, in addition to the intravalley peak at low momenta. The ellipsoidal structure of the scattering maxima is a consequence of the anisotropic dispersion at the $M$-point, as confirmed by both experimental data and DFT calculations (Extended Data Fig. 5). 

We note that the shape of the potential perturbation caused by the SV is likely determined by the lattice symmetry and thus posit that each SV forms a triangular potential perturbation extending towards the neighboring Pt atoms. The magnitudes of this spatially extended perturbation likely depend on the distance from the vacancy, with the largest perturbation being at the vacancy center. When fitting for the values of potential hills, we find that having two parameters is sufficient to accurately reproduce the small-scale variation in the SV spatial map and FFT result (Fig.~\ref{fig:SV}f,h). This validates our characterization of QADs as a framework made up of SVs, where each SV contributes a potential hill to scatter the quasiparticles in PtTe$_2$. The origin of the fine-scale variation in the quantum states lies in the intervalley scattering between $M$-points, where the large in-plane momenta result in short oscillatory wavelengths. 

An SV has no prominent bound state but once SVs aggregate close to each other to form a QAD, the perturbation is large enough for discrete energy levels to form. Here, we mainly focus on the discussion of the $dI/dV$ features for QADs inside the gap, which are associated with antidot bound states. The formation of these energy levels likely involves the states from both the BB and TB. Given the complexity of the band structure, we take a pragmatic approach and use three independent $\mathbf{k}\cdot\mathbf{p}$ expansions to describe the $M$-point BBM and the two TBM's at $\Gamma$- and $K$-points. The effective anisotropic masses are obtained from DFT calculations. As mentioned above, we simulate QADs with SVs as the building blocks and fit the parameters of the potential perturbation only to the SV. As a result, in Fig.~\ref{fig:QAD}(u-x), our calculation of the spectral functions obtain the resonance peaks for trimer at 2.46 eV, hexamer at 2.32 eV, 2.70 eV and 2.90 eV, decamer at 2.35 eV, 2.60 eV, 2.71 eV and 3.00 eV, pentadecamer at 2.30 eV, 2.55 eV, 2.67 eV, 2.84 eV and 3.07 eV with the BBM and TBM band edges at 2.3 eV and 2.75 eV, respectively. The larger the QAD is, the more resonance peaks with smaller energy spacing are obtained, which show good agreement with experimental data. The small variation of the energy position between the experimental data and simulation may come from the increasing repulsive potential from trimer to pentadecamer as measured by KPFM (Extended Data Fig. 6). Our model also suggests that the quantum states form as a result from scattering not only from the same band, but also from different bands. To leading order, the combined spectral function from the BBM and TBM is a linear superposition of spectral functions from individual bands. However, the values of the potential perturbation are likely to be different across the two bands. The band structure composition from DFT calculations in Extended Data Fig. 5 show that the BB and TB are composed of Te states and both Pt and Te states, respectively. This means that the BB is likely to be more affected by the Te vacancy than the TB, leading to a larger overall contribution to the spectral function from the BB scattering. We consider a linear combination of the spectral function from both the BB and TB, with the BB having a higher contribution to the total spectral function, as shown in Extended Data Fig. 10. The simulation results reproduce the spatial distribution of all the quantum states observed experimentally (Fig.~\ref{fig:QAD}(e-t) and Extended Data Fig. 7). The $P_1$ and $P_2$ peaks can be understood as ``bonding state'' with s-orbital-like pattern and ``antibonding state'' with a more complex nodal pattern. These two peaks are in a reversal energy sequence because holes at is in lower energy at $P_1$ than $P_2$.

In addition, we investigated the influence of oxygen substitution in QADs. DFT calculations reveal that both substitutional O and O-O can bond with the adjacent Pt atoms next to the vacancy site, which exhibit a similar DOS feature as pristine PtTe$_2$ (Extended Data Fig. 5). Hence, the substitution of O and O-O here can be seen as a passivation of Te vacancies with respect to the electronic structure of PtTe$_2$. However, the O-O substitutes still works as a scattering center that maintains the geometry of the QAD but with a slightly altered local potential perturbation compared to Te vacancies. After fitting the parameters to account for the changed defect species in our model, the spatial distribution of the quantum states are well reproduced from 5V1O$_2$ to 3V3O$_2$ (Fig.~\ref{fig:Oxygen}(n-p)). Hence, O-O substitute still behaves as a repulsive potential to the electrons and maintains the size and shape of the QAD. No matter how many vacancies are substituted, the oxygen doped QADs still hold the same quantum states with slightly tunable energy positions depending on the doping level, which suggests a remarkable robustness of these quantum states. Besides the O-O substitute, vacancies can act as a template to hold foreign atoms with designed properties. Examples include introducing N or P atom to form attractive potential centers to result in quantum hetero-antidots, and introducing spin- polarized atoms to explore Skew scattering and their potential applications in QAD-based spintronics.

\subsection{Conclusion}
In summary, we demonstrated a bottom-up self-assembly of Te SVs on PtTe$_2$ surface into atomically-precise QADs, hosting precisely engineered multi-level quantum hole states inside QADs with energy spacing widely tunable from telecom to far-infrared regime. Inter- and intra-valley scattering at the $M$-pocket dominates electronic scattering, enabling the modulation of the nodal pattern of these bound states at the extraordinary fine scale. In addition, SV-based QADs are symmetry-protected and thus survive upon oxygen substitution of Te vacancies, indicating remarkable robustness and tunability in quantum properties, crucial for their practical applications. The ability to assemble such atomic Legos into QADs at a level of atomic precision unobtainable by conventional top-down lithography with molecule-sized discrete quantum states without a magnetic field is central to the realization of QADs-based devices with atom precision. Further modulation of these QADs by introducing spin-polarized atoms to fabricate magnetic QADs and antiferromagnetic Ising systems on a triangular lattice may help to gain atomic insight into skew scattering~\citep{Ishizuka2018,Fujishiro2021} and exotic quantum phases~\citep{Arh2022}.

\bibliography{ptte2}

\begin{thebibliography}{10}
\expandafter\ifx\csname url\endcsname\relax
  \def\url#1{\texttt{#1}}\fi
\expandafter\ifx\csname urlprefix\endcsname\relax\def\urlprefix{URL }\fi
\providecommand{\bibinfo}[2]{#2}
\providecommand{\eprint}[2][]{\url{#2}}

\bibitem{Weiss1991}
\bibinfo{author}{Weiss, D.} \emph{et~al.}
\newblock \bibinfo{title}{Electron pinball and commensurate orbits in a
  periodic array of scatterers}.
\newblock \emph{\bibinfo{journal}{Phys. Rev. Lett.}}
  \textbf{\bibinfo{volume}{66}}, \bibinfo{pages}{2790--2793}
  (\bibinfo{year}{1991}).

\bibitem{Fleischmann1992}
\bibinfo{author}{Fleischmann, R.}, \bibinfo{author}{Geisel, T.} \&
  \bibinfo{author}{Ketzmerick, R.}
\newblock \bibinfo{title}{Magnetoresistance due to chaos and nonlinear
  resonances in lateral surface superlattices}.
\newblock \emph{\bibinfo{journal}{Phys. Rev. Lett.}}
  \textbf{\bibinfo{volume}{68}}, \bibinfo{pages}{1367--1370}
  (\bibinfo{year}{1992}).

\bibitem{Gunawan2008}
\bibinfo{author}{Gunawan, O.}, \bibinfo{author}{Gokmen, T.},
  \bibinfo{author}{Shkolnikov, Y.~P.}, \bibinfo{author}{De~Poortere, E.~P.} \&
  \bibinfo{author}{Shayegan, M.}
\newblock \bibinfo{title}{Anomalous giant piezoresistance in alas 2d electron
  systems with antidot lattices}.
\newblock \emph{\bibinfo{journal}{Phys. Rev. Lett.}}
  \textbf{\bibinfo{volume}{100}}, \bibinfo{pages}{036602}
  (\bibinfo{year}{2008}).

\bibitem{Ouyang2011}
\bibinfo{author}{Ouyang, F.}, \bibinfo{author}{Peng, S.}, \bibinfo{author}{Liu,
  Z.} \& \bibinfo{author}{Liu, Z.}
\newblock \bibinfo{title}{Bandgap opening in graphene antidot lattices: The
  missing half}.
\newblock \emph{\bibinfo{journal}{ACS Nano}} \textbf{\bibinfo{volume}{5}},
  \bibinfo{pages}{4023--4030} (\bibinfo{year}{2011}).

\bibitem{Cupo2017}
\bibinfo{author}{Cupo, A.} \emph{et~al.}
\newblock \bibinfo{title}{Periodic arrays of phosphorene nanopores as antidot
  lattices with tunable properties}.
\newblock \emph{\bibinfo{journal}{ACS Nano}} \textbf{\bibinfo{volume}{11}},
  \bibinfo{pages}{7494--7507} (\bibinfo{year}{2017}).

\bibitem{Du2018}
\bibinfo{author}{Du, L.} \emph{et~al.}
\newblock \bibinfo{title}{Emerging many-body effects in semiconductor
  artificial graphene with low disorder}.
\newblock \emph{\bibinfo{journal}{Nature Communications}}
  \textbf{\bibinfo{volume}{9}}, \bibinfo{pages}{3299} (\bibinfo{year}{2018}).

\bibitem{Liu2021}
\bibinfo{author}{Liu, M.}, \bibinfo{author}{Nam, H.}, \bibinfo{author}{Kim,
  J.}, \bibinfo{author}{Fiete, G.~A.} \& \bibinfo{author}{Shih, C.-K.}
\newblock \bibinfo{title}{Influence of nanosize hole defects and their
  geometric arrangements on the superfluid density in atomically thin single
  crystals of indium superconductor}.
\newblock \emph{\bibinfo{journal}{Physical Review Letters}}
  \textbf{\bibinfo{volume}{127}}, \bibinfo{pages}{127003}
  (\bibinfo{year}{2021}).

\bibitem{Sandner2015}
\bibinfo{author}{Sandner, A.} \emph{et~al.}
\newblock \bibinfo{title}{Ballistic transport in graphene antidot lattices}.
\newblock \emph{\bibinfo{journal}{Nano Letters}} \textbf{\bibinfo{volume}{15}},
  \bibinfo{pages}{8402--8406} (\bibinfo{year}{2015}).

\bibitem{Jessen2019}
\bibinfo{author}{Jessen, B.~S.} \emph{et~al.}
\newblock \bibinfo{title}{Lithographic band structure engineering of graphene}.
\newblock \emph{\bibinfo{journal}{Nature Nanotechnology}}
  \textbf{\bibinfo{volume}{14}}, \bibinfo{pages}{340--346}
  (\bibinfo{year}{2019}).

\bibitem{Du2009}
\bibinfo{author}{Du, A.} \emph{et~al.}
\newblock \bibinfo{title}{Dots versus antidots: Computational exploration of
  structure, magnetism, and half-metallicity in boron-nitride nanostructures}.
\newblock \emph{\bibinfo{journal}{Journal of the American Chemical Society}}
  \textbf{\bibinfo{volume}{131}}, \bibinfo{pages}{17354--17359}
  (\bibinfo{year}{2009}).

\bibitem{Mitterreiter2021}
\bibinfo{author}{Mitterreiter, E.} \emph{et~al.}
\newblock \bibinfo{title}{The role of chalcogen vacancies for atomic defect
  emission in mos2}.
\newblock \emph{\bibinfo{journal}{Nature Communications}}
  \textbf{\bibinfo{volume}{12}}, \bibinfo{pages}{3822} (\bibinfo{year}{2021}).

\bibitem{Flindt2005}
\bibinfo{author}{Flindt, C.}, \bibinfo{author}{Mortensen, N.~A.} \&
  \bibinfo{author}{Jauho, A.-P.}
\newblock \bibinfo{title}{Quantum computing via defect states in
  {Two-Dimensional} antidot lattices}.
\newblock \emph{\bibinfo{journal}{Nano Lett.}} \textbf{\bibinfo{volume}{5}},
  \bibinfo{pages}{2515--2518} (\bibinfo{year}{2005}).

\bibitem{Pedersen2008}
\bibinfo{author}{Pedersen, T.~G.} \emph{et~al.}
\newblock \bibinfo{title}{Graphene antidot lattices: Designed defects and spin
  qubits}.
\newblock \emph{\bibinfo{journal}{Phys. Rev. Lett.}}
  \textbf{\bibinfo{volume}{100}}, \bibinfo{pages}{136804}
  (\bibinfo{year}{2008}).

\bibitem{Besteiro2017}
\bibinfo{author}{Besteiro, L.~V.}, \bibinfo{author}{Kong, X.-T.},
  \bibinfo{author}{Wang, Z.}, \bibinfo{author}{Hartland, G.} \&
  \bibinfo{author}{Govorov, A.~O.}
\newblock \bibinfo{title}{Understanding hot-electron generation and plasmon
  relaxation in metal nanocrystals: Quantum and classical mechanisms}.
\newblock \emph{\bibinfo{journal}{ACS Photonics}} \textbf{\bibinfo{volume}{4}},
  \bibinfo{pages}{2759--2781} (\bibinfo{year}{2017}).

\bibitem{Zhang2013}
\bibinfo{author}{Zhang, H.} \emph{et~al.}
\newblock \bibinfo{title}{Large-scale mesoscopic transport in nanostructured
  graphene}.
\newblock \emph{\bibinfo{journal}{Phys. Rev. Lett.}}
  \textbf{\bibinfo{volume}{110}}, \bibinfo{pages}{066805}
  (\bibinfo{year}{2013}).

\bibitem{Clavero2014}
\bibinfo{author}{Clavero, C.}
\newblock \bibinfo{title}{Plasmon-induced hot-electron generation at
  nanoparticle/metal-oxide interfaces for photovoltaic and photocatalytic
  devices}.
\newblock \emph{\bibinfo{journal}{Nature Photonics}}
  \textbf{\bibinfo{volume}{8}}, \bibinfo{pages}{95--103}
  (\bibinfo{year}{2014}).

\bibitem{Goldman1995}
\bibinfo{author}{Goldman, V.~J.} \& \bibinfo{author}{Su, B.}
\newblock \bibinfo{title}{Resonant tunneling in the quantum hall regime:
  Measurement of fractional charge}.
\newblock \emph{\bibinfo{journal}{Science}} \textbf{\bibinfo{volume}{267}},
  \bibinfo{pages}{1010--1012} (\bibinfo{year}{1995}).

\bibitem{Maasilta2000}
\bibinfo{author}{Maasilta, I.~J.} \& \bibinfo{author}{Goldman, V.~J.}
\newblock \bibinfo{title}{Tunneling through a coherent ``quantum antidot
  molecule''}.
\newblock \emph{\bibinfo{journal}{Phys. Rev. Lett.}}
  \textbf{\bibinfo{volume}{84}}, \bibinfo{pages}{1776--1779}
  (\bibinfo{year}{2000}).

\bibitem{Sim2003}
\bibinfo{author}{Sim, H.-S.} \emph{et~al.}
\newblock \bibinfo{title}{Coulomb blockade and kondo effect in a quantum hall
  antidot}.
\newblock \emph{\bibinfo{journal}{Phys. Rev. Lett.}}
  \textbf{\bibinfo{volume}{91}}, \bibinfo{pages}{266801}
  (\bibinfo{year}{2003}).

\bibitem{Yan2017}
\bibinfo{author}{Yan, M.} \emph{et~al.}
\newblock \bibinfo{title}{Lorentz-violating type-ii dirac fermions in
  transition metal dichalcogenide ptte2}.
\newblock \emph{\bibinfo{journal}{Nature Communications}}
  \textbf{\bibinfo{volume}{8}}, \bibinfo{pages}{257} (\bibinfo{year}{2017}).

\bibitem{Fuechsle2012}
\bibinfo{author}{Fuechsle, M.} \emph{et~al.}
\newblock \bibinfo{title}{A single-atom transistor}.
\newblock \emph{\bibinfo{journal}{Nature Nanotechnology}}
  \textbf{\bibinfo{volume}{7}}, \bibinfo{pages}{242--246}
  (\bibinfo{year}{2012}).

\bibitem{Huff2018}
\bibinfo{author}{Huff, T.} \emph{et~al.}
\newblock \bibinfo{title}{Binary atomic silicon logic}.
\newblock \emph{\bibinfo{journal}{Nature Electronics}}
  \textbf{\bibinfo{volume}{1}}, \bibinfo{pages}{636--643}
  (\bibinfo{year}{2018}).

\bibitem{Achal2018}
\bibinfo{author}{Achal, R.} \emph{et~al.}
\newblock \bibinfo{title}{Lithography for robust and editable atomic-scale
  silicon devices and memories}.
\newblock \emph{\bibinfo{journal}{Nature Communications}}
  \textbf{\bibinfo{volume}{9}}, \bibinfo{pages}{2778} (\bibinfo{year}{2018}).

\bibitem{Kalff2016}
\bibinfo{author}{Kalff, F.~E.} \emph{et~al.}
\newblock \bibinfo{title}{A kilobyte rewritable atomic memory}.
\newblock \emph{\bibinfo{journal}{Nature Nanotechnology}}
  \textbf{\bibinfo{volume}{11}}, \bibinfo{pages}{926--929}
  (\bibinfo{year}{2016}).

\bibitem{Amlani1999}
\bibinfo{author}{Amlani, I.} \emph{et~al.}
\newblock \bibinfo{title}{Digital logic gate using quantum-dot cellular
  automata}.
\newblock \emph{\bibinfo{journal}{Science}} \textbf{\bibinfo{volume}{284}},
  \bibinfo{pages}{289--291} (\bibinfo{year}{1999}).

\bibitem{Imre2006}
\bibinfo{author}{Imre, A.} \emph{et~al.}
\newblock \bibinfo{title}{Majority logic gate for magnetic quantum-dot cellular
  automata}.
\newblock \emph{\bibinfo{journal}{Science}} \textbf{\bibinfo{volume}{311}},
  \bibinfo{pages}{205--208} (\bibinfo{year}{2006}).

\bibitem{Kim2014}
\bibinfo{author}{Kim, D.} \emph{et~al.}
\newblock \bibinfo{title}{Quantum control and process tomography of a
  semiconductor quantum dot hybrid qubit}.
\newblock \emph{\bibinfo{journal}{Nature}} \textbf{\bibinfo{volume}{511}},
  \bibinfo{pages}{70--74} (\bibinfo{year}{2014}).

\bibitem{Folsch2014}
\bibinfo{author}{F{\"o}lsch, S.}, \bibinfo{author}{Mart{\'i}nez-Blanco, J.},
  \bibinfo{author}{Yang, J.}, \bibinfo{author}{Kanisawa, K.} \&
  \bibinfo{author}{Erwin, S.~C.}
\newblock \bibinfo{title}{Quantum dots with single-atom precision}.
\newblock \emph{\bibinfo{journal}{Nature Nanotechnology}}
  \textbf{\bibinfo{volume}{9}}, \bibinfo{pages}{505--508}
  (\bibinfo{year}{2014}).

\bibitem{Park1997}
\bibinfo{author}{Park, M.}, \bibinfo{author}{Harrison, C.},
  \bibinfo{author}{Chaikin, P.~M.}, \bibinfo{author}{Register, R.~A.} \&
  \bibinfo{author}{Adamson, D.~H.}
\newblock \bibinfo{title}{Block copolymer lithography: Periodic arrays of
  {\textasciitilde}1011 holes in 1 square centimeter}.
\newblock \emph{\bibinfo{journal}{Science}} \textbf{\bibinfo{volume}{276}},
  \bibinfo{pages}{1401--1404} (\bibinfo{year}{1997}).

\bibitem{Sinitskii2010}
\bibinfo{author}{Sinitskii, A.} \& \bibinfo{author}{Tour, J.~M.}
\newblock \bibinfo{title}{Patterning graphene through the self-assembled
  templates: Toward periodic two-dimensional graphene nanostructures with
  semiconductor properties}.
\newblock \emph{\bibinfo{journal}{Journal of the American Chemical Society}}
  \textbf{\bibinfo{volume}{132}}, \bibinfo{pages}{14730--14732}
  (\bibinfo{year}{2010}).

\bibitem{Khajetoorians2019}
\bibinfo{author}{Khajetoorians, A.~A.}, \bibinfo{author}{Wegner, D.},
  \bibinfo{author}{Otte, A.~F.} \& \bibinfo{author}{Swart, I.}
\newblock \bibinfo{title}{Creating designer quantum states of matter
  atom-by-atom}.
\newblock \emph{\bibinfo{journal}{Nature Reviews Physics}}
  \textbf{\bibinfo{volume}{1}}, \bibinfo{pages}{703--715}
  (\bibinfo{year}{2019}).

\bibitem{Gomes2012}
\bibinfo{author}{Gomes, K.~K.}, \bibinfo{author}{Mar, W.}, \bibinfo{author}{Ko,
  W.}, \bibinfo{author}{Guinea, F.} \& \bibinfo{author}{Manoharan, H.~C.}
\newblock \bibinfo{title}{Designer dirac fermions and topological phases in
  molecular graphene}.
\newblock \emph{\bibinfo{journal}{Nature}} \textbf{\bibinfo{volume}{483}},
  \bibinfo{pages}{306--310} (\bibinfo{year}{2012}).

\bibitem{Slot2017}
\bibinfo{author}{Slot, M.~R.} \emph{et~al.}
\newblock \bibinfo{title}{Experimental realization and characterization of an
  electronic lieb lattice}.
\newblock \emph{\bibinfo{journal}{Nature Physics}}
  \textbf{\bibinfo{volume}{13}}, \bibinfo{pages}{672--676}
  (\bibinfo{year}{2017}).

\bibitem{Kempkes2019a}
\bibinfo{author}{Kempkes, S.~N.} \emph{et~al.}
\newblock \bibinfo{title}{Design and characterization of electrons in a fractal
  geometry}.
\newblock \emph{\bibinfo{journal}{Nature Physics}}
  \textbf{\bibinfo{volume}{15}}, \bibinfo{pages}{127--131}
  (\bibinfo{year}{2019}).

\bibitem{Drost2017}
\bibinfo{author}{Drost, R.}, \bibinfo{author}{Ojanen, T.},
  \bibinfo{author}{Harju, A.} \& \bibinfo{author}{Liljeroth, P.}
\newblock \bibinfo{title}{Topological states in engineered atomic lattices}.
\newblock \emph{\bibinfo{journal}{Nature Physics}}
  \textbf{\bibinfo{volume}{13}}, \bibinfo{pages}{668--671}
  (\bibinfo{year}{2017}).

\bibitem{Xu2002}
\bibinfo{author}{Xu, J.} \emph{et~al.}
\newblock \bibinfo{title}{Quantum antidot formation and correlation to optical
  shift of gold nanoparticles embedded in mgo}.
\newblock \emph{\bibinfo{journal}{Phys. Rev. Lett.}}
  \textbf{\bibinfo{volume}{88}}, \bibinfo{pages}{175502}
  (\bibinfo{year}{2002}).

\bibitem{Liu2014}
\bibinfo{author}{Liu, Y.}, \bibinfo{author}{Xu, F.}, \bibinfo{author}{Zhang,
  Z.}, \bibinfo{author}{Penev, E.~S.} \& \bibinfo{author}{Yakobson, B.~I.}
\newblock \bibinfo{title}{Two-dimensional mono-elemental semiconductor with
  electronically inactive defects: The case of phosphorus}.
\newblock \emph{\bibinfo{journal}{Nano Letters}} \textbf{\bibinfo{volume}{14}},
  \bibinfo{pages}{6782--6786} (\bibinfo{year}{2014}).

\bibitem{Nguyen2018}
\bibinfo{author}{Nguyen, G.~D.} \emph{et~al.}
\newblock \bibinfo{title}{3d imaging and manipulation of subsurface selenium
  vacancies in ${\mathrm{pdse}}_{2}$}.
\newblock \emph{\bibinfo{journal}{Phys. Rev. Lett.}}
  \textbf{\bibinfo{volume}{121}}, \bibinfo{pages}{086101}
  (\bibinfo{year}{2018}).

\bibitem{LiXZ2021}
\bibinfo{author}{Li, X.} \emph{et~al.}
\newblock \bibinfo{title}{Ordered clustering of single atomic te vacancies in
  atomically thin ptte2 promotes hydrogen evolution catalysis}.
\newblock \emph{\bibinfo{journal}{Nature Communications}}
  \textbf{\bibinfo{volume}{12}}, \bibinfo{pages}{2351} (\bibinfo{year}{2021}).

\bibitem{Zhussupbekov2021}
\bibinfo{author}{Zhussupbekov, K.} \emph{et~al.}
\newblock \bibinfo{title}{Imaging and identification of point defects in
  ptte2}.
\newblock \emph{\bibinfo{journal}{npj 2D Materials and Applications}}
  \textbf{\bibinfo{volume}{5}}, \bibinfo{pages}{14} (\bibinfo{year}{2021}).

\bibitem{Leo2009}
\bibinfo{author}{Leo, G.}, \bibinfo{author}{Fabian, M.},
  \bibinfo{author}{Nikolaj, M.}, \bibinfo{author}{Peter, L.} \&
  \bibinfo{author}{Gerhard, M.}
\newblock \bibinfo{title}{The chemical structure of a molecule resolved by
  atomic force microscopy}.
\newblock \emph{\bibinfo{journal}{Science}} \textbf{\bibinfo{volume}{325}},
  \bibinfo{pages}{1110--1114} (\bibinfo{year}{2009}).

\bibitem{Barja2019}
\bibinfo{author}{Barja, S.} \emph{et~al.}
\newblock \bibinfo{title}{Identifying substitutional oxygen as a prolific point
  defect in monolayer transition metal dichalcogenides}.
\newblock \emph{\bibinfo{journal}{Nature Communications}}
  \textbf{\bibinfo{volume}{10}}, \bibinfo{pages}{3382} (\bibinfo{year}{2019}).

\bibitem{Schuler2019ACSnano}
\bibinfo{author}{Schuler, B.} \emph{et~al.}
\newblock \bibinfo{title}{How substitutional point defects in two-dimensional
  ws2 induce charge localization, spin--orbit splitting, and strain}.
\newblock \emph{\bibinfo{journal}{ACS Nano}} \textbf{\bibinfo{volume}{13}},
  \bibinfo{pages}{10520--10534} (\bibinfo{year}{2019}).

\bibitem{Cochrane2021}
\bibinfo{author}{Cochrane, K.~A.} \emph{et~al.}
\newblock \bibinfo{title}{Spin-dependent vibronic response of a carbon radical
  ion in two-dimensional ws2}.
\newblock \emph{\bibinfo{journal}{Nature Communications}}
  \textbf{\bibinfo{volume}{12}}, \bibinfo{pages}{7287} (\bibinfo{year}{2021}).

\bibitem{Guo1986}
\bibinfo{author}{Guo, G.~Y.} \& \bibinfo{author}{Liang, W.~Y.}
\newblock \bibinfo{title}{The electronic structures of platinum
  dichalcogenides: Pts2, ptse2 and ptte2}.
\newblock \emph{\bibinfo{journal}{Journal of Physics C: Solid State Physics}}
  \textbf{\bibinfo{volume}{19}}, \bibinfo{pages}{995} (\bibinfo{year}{1986}).

\bibitem{Aghajanian2020}
\bibinfo{author}{Aghajanian, M.} \emph{et~al.}
\newblock \bibinfo{title}{Resonant and bound states of charged defects in
  two-dimensional semiconductors}.
\newblock \emph{\bibinfo{journal}{Phys. Rev. B}}
  \textbf{\bibinfo{volume}{101}}, \bibinfo{pages}{081201}
  (\bibinfo{year}{2020}).

\bibitem{Fang2022}
\bibinfo{author}{Fang, H.} \emph{et~al.}
\newblock \bibinfo{title}{Electronic self-passivation of single vacancy in
  black phosphorus via ionization}.
\newblock \emph{\bibinfo{journal}{Phys. Rev. Lett.}}
  \textbf{\bibinfo{volume}{128}}, \bibinfo{pages}{176801}
  (\bibinfo{year}{2022}).

\bibitem{Schuler2019PRL}
\bibinfo{author}{Schuler, B.} \emph{et~al.}
\newblock \bibinfo{title}{Large spin-orbit splitting of deep in-gap defect
  states of engineered sulfur vacancies in monolayer ${\mathrm{ws}}_{2}$}.
\newblock \emph{\bibinfo{journal}{Phys. Rev. Lett.}}
  \textbf{\bibinfo{volume}{123}}, \bibinfo{pages}{076801}
  (\bibinfo{year}{2019}).

\bibitem{Gross2014}
\bibinfo{author}{Gross, L.} \emph{et~al.}
\newblock \bibinfo{title}{Investigating atomic contrast in atomic force
  microscopy and kelvin probe force microscopy on ionic systems using
  functionalized tips}.
\newblock \emph{\bibinfo{journal}{Physical Review B}}
  \textbf{\bibinfo{volume}{90}}, \bibinfo{pages}{155455}
  (\bibinfo{year}{2014}).

\bibitem{Cai2016}
\bibinfo{author}{Cai, Y.}, \bibinfo{author}{Ke, Q.}, \bibinfo{author}{Zhang,
  G.}, \bibinfo{author}{Yakobson, B.~I.} \& \bibinfo{author}{Zhang, Y.-W.}
\newblock \bibinfo{title}{Highly itinerant atomic vacancies in phosphorene}.
\newblock \emph{\bibinfo{journal}{Journal of the American Chemical Society}}
  \textbf{\bibinfo{volume}{138}}, \bibinfo{pages}{10199--10206}
  (\bibinfo{year}{2016}).

\bibitem{Trevethan2014}
\bibinfo{author}{Trevethan, T.}, \bibinfo{author}{Latham, C.~D.},
  \bibinfo{author}{Heggie, M.~I.}, \bibinfo{author}{Briddon, P.~R.} \&
  \bibinfo{author}{Rayson, M.~J.}
\newblock \bibinfo{title}{Vacancy diffusion and coalescence in graphene
  directed by defect strain fields}.
\newblock \emph{\bibinfo{journal}{Nanoscale}} \textbf{\bibinfo{volume}{6}},
  \bibinfo{pages}{2978--2986} (\bibinfo{year}{2014}).

\bibitem{Ishizuka2018}
\bibinfo{author}{Ishizuka, H.} \& \bibinfo{author}{Nagaosa, N.}
\newblock \bibinfo{title}{Spin chirality induced skew scattering and anomalous
  hall effect in chiral magnets}.
\newblock \emph{\bibinfo{journal}{Science Advances}}
  \textbf{\bibinfo{volume}{4}}, \bibinfo{pages}{eaap9962}
  (\bibinfo{year}{2018}).

\bibitem{Fujishiro2021}
\bibinfo{author}{Fujishiro, Y.} \emph{et~al.}
\newblock \bibinfo{title}{Giant anomalous hall effect from spin-chirality
  scattering in a chiral magnet}.
\newblock \emph{\bibinfo{journal}{Nature Communications}}
  \textbf{\bibinfo{volume}{12}}, \bibinfo{pages}{317} (\bibinfo{year}{2021}).

\bibitem{Arh2022}
\bibinfo{author}{Arh, T.} \emph{et~al.}
\newblock \bibinfo{title}{The ising triangular-lattice antiferromagnet
  neodymium heptatantalate as a quantum spin liquid candidate}.
\newblock \emph{\bibinfo{journal}{Nature Materials}}
  \textbf{\bibinfo{volume}{21}}, \bibinfo{pages}{416--422}
  (\bibinfo{year}{2022}).

\bibitem{Julia-2017}
\bibinfo{author}{Bezanson, J.}, \bibinfo{author}{Edelman, A.},
  \bibinfo{author}{Karpinski, S.} \& \bibinfo{author}{Shah, V.~B.}
\newblock \bibinfo{title}{Julia: A fresh approach to numerical computing}.
\newblock \emph{\bibinfo{journal}{SIAM {R}eview}}
  \textbf{\bibinfo{volume}{59}}, \bibinfo{pages}{65--98}
  (\bibinfo{year}{2017}).

\bibitem{Kresse1996}
\bibinfo{author}{Kresse, G.} \& \bibinfo{author}{Furthm\"uller, J.}
\newblock \bibinfo{title}{Efficient iterative schemes for ab initio
  total-energy calculations using a plane-wave basis set}.
\newblock \emph{\bibinfo{journal}{Phys. Rev. B}} \textbf{\bibinfo{volume}{54}},
  \bibinfo{pages}{11169--11186} (\bibinfo{year}{1996}).

\bibitem{Blochl1994}
\bibinfo{author}{Bl\"ochl, P.~E.}
\newblock \bibinfo{title}{Projector augmented-wave method}.
\newblock \emph{\bibinfo{journal}{Phys. Rev. B}} \textbf{\bibinfo{volume}{50}},
  \bibinfo{pages}{17953--17979} (\bibinfo{year}{1994}).

\bibitem{Perdew1996}
\bibinfo{author}{Perdew, J.~P.}, \bibinfo{author}{Burke, K.} \&
  \bibinfo{author}{Ernzerhof, M.}
\newblock \bibinfo{title}{Generalized gradient approximation made simple}.
\newblock \emph{\bibinfo{journal}{Phys. Rev. Lett.}}
  \textbf{\bibinfo{volume}{77}}, \bibinfo{pages}{3865--3868}
  (\bibinfo{year}{1996}).

\bibitem{Moellmann2014}
\bibinfo{author}{Moellmann, J.} \& \bibinfo{author}{Grimme, S.}
\newblock \bibinfo{title}{Dft-d3 study of some molecular crystals}.
\newblock \emph{\bibinfo{journal}{The Journal of Physical Chemistry C}}
  \textbf{\bibinfo{volume}{118}}, \bibinfo{pages}{7615--7621}
  (\bibinfo{year}{2014}).

\bibitem{Henkelman2000}
\bibinfo{author}{Henkelman, G.}, \bibinfo{author}{Uberuaga, B.~P.} \&
  \bibinfo{author}{Jónsson, H.}
\newblock \bibinfo{title}{A climbing image nudged elastic band method for
  finding saddle points and minimum energy paths}.
\newblock \emph{\bibinfo{journal}{The Journal of Chemical Physics}}
  \textbf{\bibinfo{volume}{113}}, \bibinfo{pages}{9901--9904}
  (\bibinfo{year}{2000}).

\bibitem{Giannozzi2009}
\bibinfo{author}{Giannozzi, P.} \emph{et~al.}
\newblock \bibinfo{title}{{QUANTUM ESPRESSO: A modular and open-source software
  project for quantum simulations of materials}}.
\newblock \emph{\bibinfo{journal}{Journal of Physics Condensed Matter}}
  \textbf{\bibinfo{volume}{21}}, \bibinfo{pages}{395502}
  (\bibinfo{year}{2009}).

\bibitem{Giannozzi2017}
\bibinfo{author}{Giannozzi, P.} \emph{et~al.}
\newblock \bibinfo{title}{{Advanced capabilities for materials modelling with
  Quantum ESPRESSO}}.
\newblock \emph{\bibinfo{journal}{Journal of Physics Condensed Matter}}
  \textbf{\bibinfo{volume}{29}}, \bibinfo{pages}{465901}
  (\bibinfo{year}{2017}).

\bibitem{DalCorso2014}
\bibinfo{author}{{Dal Corso}, A.}
\newblock \bibinfo{title}{{Pseudopotentials periodic table: From H to Pu}}.
\newblock \emph{\bibinfo{journal}{Computational Materials Science}}
  \textbf{\bibinfo{volume}{95}}, \bibinfo{pages}{337--350}
  (\bibinfo{year}{2014}).

\end{thebibliography}

\clearpage

\section{Methods}

\subsection{Crystal synthesis}
\noindent
PtTe$_2$ single crystals were synthesized in a three-zone tube furnace (OTF-1200X-III-S-UL, MTI Corporation, USA) using the previously reported chemical vapour transport method~\citep{LiXZ2021}. Firstly, 1 g of high-purity Pt (0.433 g) and Te (0.567 g) powders with a stoichiometric molar ratio of 1:2 were mixed thoroughly in a glove box. The mixed powders were then sealed in an evacuated quartz tube (length: 10 cm; external diameter: 13 mm; and wall thickness: 1 mm) and heated up to 1000 $^{\circ}$C for 48 h in the furnace. The temperature was further increased to 1150 $^{\circ}$C for another 1 h. Finally, the furnace was slowly cooled down to ambient temperature at a speed of 5 $^{\circ}$C/min for collecting single crystals.

\subsection{STM/nc-AFM characterization}
\noindent
The PtTe$_2$ bulk sample was of 2-3 mm thick and was securely enveloped by Tantalum (Ta) foil and fixed on the sample plate by spot welding. PtTe$_2$ bulk sample was mechanically exfoliated by Scotch tape method to remove the top layers and thus to expose fresh clean surface inside the load lock of Createc LT-STM/nc-AFM system under vacuum condition ($~10^{-7}$ mbar). The sample was then transferred to the scanning head for further measurement at low temperature ($T=4.8$K) and ultrahigh vacuum ($<10^{-10}$ mbar). The STM tip was calibrated on the Au (111) surface by checking the Shockley surface state. All the $dI/dV$ spectra were taken through the standard lock-in technique with a modulation voltage of 10 mV and frequency of 877 Hz. All the spectra were averaged from the measurement taken at least three different QADs to rule out the influence of random potential deviation caused by the nearby charged features. The nc-AFM was conducted with a commercial qPlus sensor. The resonant frequency of the tuning fork was $f_0 = 28$ kHz, the stiffness was $k = 1800$ NM$^{-1}$, and the quality factor was above 16k. nc-AFM imaging with the qPlus sensor was performed at the frequency-modulation mode with an oscillation amplitude of $A = 50$ pm. The STM/nc-AFM images were collected in constant height mode simultaneously with a CO functionalized tip.

\subsection{QFT calculations}
\noindent
The derivation of the field-theoretic formalism is detailed in the Supplementary Information. We use a single-band nearly-free electron model with direction-dependent dispersion to describe the surface states of PtTe$_2$. The triangular potential perturbation caused by a single Te vacancy is parametrized by two values of potential hill height, $U_1 = 0.22$V and $U_2 = 0.155$V, with $U_1$ located at the location of the Te vacancy. Similarly, the potential perturbation caused by the O substitution is parametrized by two values of potential hill height, $U_3 = 0.14$V and $U_4 = 0.12$V, with $U_3$ located at the O atom. 
The numerical calculations were performed using the \textsc{JULIA} programming language~\citep{Julia-2017}.

\subsection{Diffusion barrier calculations}
\noindent
The diffusion barrier calculations for PtTe$_2$ were based on the density functional theory as implemented in the Vienna Ab initio Simulation Package (VASP)~\citep{Kresse1996}. The projector augmented-wave (PAW) method~\citep{Blochl1994} and the Perdew-Burke-Ernzerhof type generalized gradient approximation (GGA-PBE)~\citep{Perdew1996} were performed to represent electron-ion interactions and the exchange-correlation interactions, respectively. The van der Waals (vdW) correction with the DFT-D3 method was adopted~\citep{Moellmann2014} and a vacuum slab large enough $\sim 18$\AA was applied in the z direction to prevent interaction between two neighboring PtTe$_2$ layers. All the structures were optimized until the forces on each atom were smaller than 0.01 $eV/$\AA\ and the energy convergence criteria was set to 10-4 eV. The Brillouin zone was sample at the $\Gamma$-Point and the energy cutoff for the plane-wave basis set was chosen to be 400 eV. The activation barrier associated with Te vacancy migration was derived using climbing image nudged-elastic band (CL-NEB) method~\citep{Henkelman2000}.

\subsection{DFT band structure calculations}
\noindent
Density function theory band structure calculations were performed with Quantum ESPRESSO~\citep{Giannozzi2009,Giannozzi2017} using a projector augmented wave (PAW) basis~\citep{Blochl1994,DalCorso2014} and the Perdew-Burke-Ernzerhof (PBE)~\citep{Perdew1996} exchange correlation functional. The kinetic energy cutoff of wave functions was set to the minimum recommended value specified by the basis~\citep{DalCorso2014}. Spin orbit corrections were included. For charge density calculations, the Brillouin zone was sampled using a uniform $18 \times 18 \times 24$ grid.

\medskip
\noindent \textbf{Acknowledgements}
J.Lu acknowledges the support from MOE grants (MOE2019-T2-2-044, MOE-T2EP50121-0008, MOE-T2EP10221-0005) and MOE (Singapore) through the Research Centre of Excellence program (Grant No. EDUN C-33-18-279-V12, I-FIM), and Agency for Science, Technology and Research (A*STAR) under its AME IRG Grant (Project No. M21K2c0113). A.R. acknowledges the National Research Foundation, Prime Minister Office, Singapore, under its Medium Sized Centre Programme and the support by Yale-NUS College (through Grant No. R-607-265-380-121).

\medskip
\noindent \textbf{Author contributions}
J.Lu supervised the project and organized the collaboration. H.F. and J. Lu conceived and designed the experiments H.F. carried out all the STM/nc-AFM measurements. Z.Q., Y.H. and T.Y. assisted the measurements. A.R, H.M. and D.D. conceived theoretical studies and carried out the quantum field theoretical calculations. X.L. and C.S. prepared as-designed PtTe$_2$ single crystal. X.H. and K.N. performed the DFT calculations. H.C. and Y.C. carried out the vacancy diffusion-related calculations. P.L., J.Li and W.C. contributed to the scientific discussion and paper preparation. H.F., H.M., A.R. and J.Lu analysed the data and wrote the paper with inputs from all authors.

\medskip
\noindent \textbf{Competing financial interests}
The authors declare no competing financial interests.

\clearpage

\end{document}